\documentclass{ws-p8-50x6-00}

\def\lsim{\mathrel{\rlap{\lower4pt\hbox{\hskip1pt$\sim$}}
    \raise1pt\hbox{$<$}}}                
\def\gsim{\mathrel{\rlap{\lower4pt\hbox{\hskip1pt$\sim$}}
    \raise1pt\hbox{$>$}}}                

\newcommand{\as}{\alpha_{\mathrm{s}}}

\newcommand{\Pmax}{\bar{q}}
\newcommand{\kt}{k_{t}}

\newcommand{\CCFM}{CCFMa,CCFMb,CCFMc,CCFMd}

\def\CASCADE{{\sc Cascade}}

\def\EPJPSI{{\sc Epjpsi}}
\def\PYTHIA{{\sc Pythia}}

\def\lsim{\mathrel{\rlap{\lower4pt\hbox{\hskip1pt$\sim$}}
    \raise1pt\hbox{$<$}}}                
\def\gsim{\mathrel{\rlap{\lower4pt\hbox{\hskip1pt$\sim$}}
    \raise1pt\hbox{$>$}}}                

\begin{document}


\title{Heavy quark production at HERA in \boldmath$k_t$ factorization
supplemented with CCFM evolution }

\author{H. Jung}

\address{   Physics Department, Lund University, 
Box 118,\\ S-221~00 Lund, Sweden\\E-mail: Hannes.Jung@desy.de }


\maketitle\abstracts{The application of $k_t$ - factorization, supplemented with the
CCFM small-$x$ evolution equation, to heavy quark production  
is discussed. Differential cross sections of $b\bar{b}$ production and
also inelastic $J/\psi$ production  
as measured at HERA are compared to the hadron level 
CCFM Monte Carlo generator \CASCADE , using the unintegrated gluon density 
obtained
within the CCFM evolution approach from a fit to HERA $F_2$ data. 
}
\section{Introduction}
The calculation of inclusive quantities, like the structure function
$F_2(x,Q^2)$ at HERA, performed in NLO QCD is in perfect agreement with the
measurements. 
But in exclusive quantities like jet or heavy quark
production large so-called $K$-factors (normalization factors)
\cite{CDF_bbar,D0_bbar}
are needed to bring the NLO calculations close to the data.
\par 
At large energies (small $x$), the evolution of parton densities proceeds over a large
region in rapidity $\Delta y \sim \log(1/x)$ and effects of finite transverse
momenta of the partons may become increasingly important.
Cross sections can then be $k_t$ - factorized\cite{CCH}
into an off-shell ($k_t$ dependent) partonic cross section
$\hat{\sigma}(\frac{x}{z},k_t) $
and a $k_t$ - unintegrated parton density function 
${\cal F}(z,k_t)$:
\begin{equation}
 \sigma  = \int 
\frac{dz}{z} d^2k_t \hat{\sigma}(\frac{x}{z},k_t) {\cal F}(z,k_t)
\label{kt-factorisation}
\end{equation}
The unintegrated gluon density ${\cal F}(z,k_t)$ is 
described by the BFKL
 evolution equation in the region of asymptotically large energies (small $x$). 
 An appropriate description valid for
both small and large $x$ is given by the CCFM evolution
equation\cite{\CCFM}, resulting in an unintegrated gluon density 
${\cal A} (x,\kt,\Pmax ) $, which is a function also of the 
additional evolution scale $\Pmax $.
\par
Catani\cite{catani-dis96,catani-feb2000} argues that
by explicitly carrying out the $k_t$ integration in eq.(\ref{kt-factorisation})
one can obtain a form fully consistent with collinear factorization: the
coefficient functions and also the DGLAP splitting functions 
are no longer evaluated in fixed order perturbation theory but
supplemented with the all-order resummation of the $\as \log 1/x$ contribution  
at small $x$. This all-loop resummation shows up in a so-called 
\emph{non-Sudakov} form factor $\Delta_{ns}$ for CCFM.
\par
In this paper
the $k_t$ factorization approach is discussed with respect to resolved
photon structure and next-to-leading corrections in the collinear approach. 
Then calculations for $b\bar{b}$ and inelastic $J/\psi$
production at HERA are presented, using 
the unintegrated gluon density obtained 
previously\cite{jung_salam_2000} from a
CCFM fit to the HERA structure function $F_2(x,Q^2)$
\section{\boldmath$k_t$-factorization versus higher order processes in collinear
factorization}
\begin{figure}[htb]
  \begin{center}
  \epsfig{figure=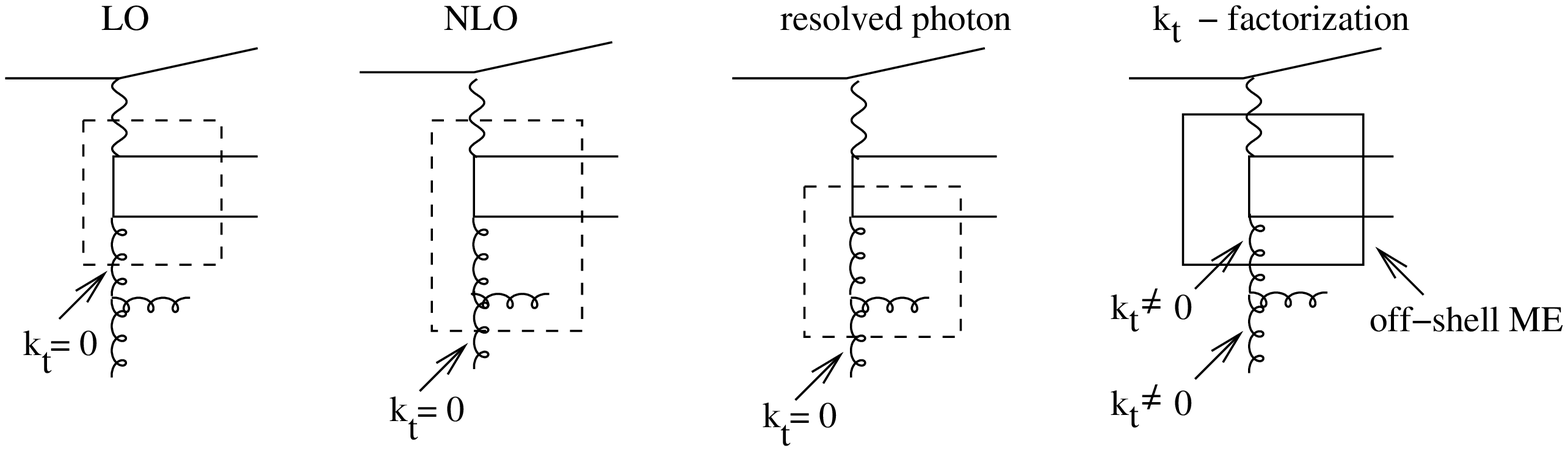,width=12cm}
  \vspace*{-0.3cm}
    \caption{{\it Diagrammatic representation of LO, NLO and resolved photon
        processes in the collinear approach compared to the
	  $k_t$-factorization approach.}}
   \label{fig:nlo-kt-fact}
  \end{center}
  \vspace*{-0.5cm}
\end{figure}
The diagrams for LO, NLO and resolved photon processes in the collinear
approach are shown schematically  in Fig.~\ref{fig:nlo-kt-fact}. In collinear
factorization, the incoming parton is on-mass-shell and has vanishing transverse
momentum $k_t$, whereas in
$k_t$-factorization the partons entering the hard scattering matrix
element are free to be off-mass-shell.  The advantage 
of the $k_t$-factorization approach becomes visible, when additional
hard gluon radiation to a $ 2 \to 2$ process like $\gamma g \to
Q\bar{Q}$ is considered. If the transverse momentum $p_{t g}$ of the
additional gluon is of the order of that of the quarks, then in the
collinear approach the full ${\cal O}(\as^2)$ matrix
element for $2 \to 3$ has to be calculated.  In $k_t$-factorization such
processes are naturally included, even if only the LO $\as$ off-shell
matrix element is used, since the $k_t$ of the incoming gluon is not
restricted from above, and therefore can acquire a virtuality similar
to the ones in a complete fixed order calculation.  In
Fig.~\ref{fig:nlo-kt-fact} the basic ideas are shown schematically .
Not only does $k_t$-factorization include (at least parts
of) NLO diagrams, it also includes diagrams of the resolved photon
type~\cite{baranov_zotov_2000},
 with the natural transition from real to virtual photons.
\par
\begin{figure}[htb]
\begin{minipage}{0.45\textwidth}
  \begin{center}
\hspace*{-2.5cm}
  \epsfig{figure=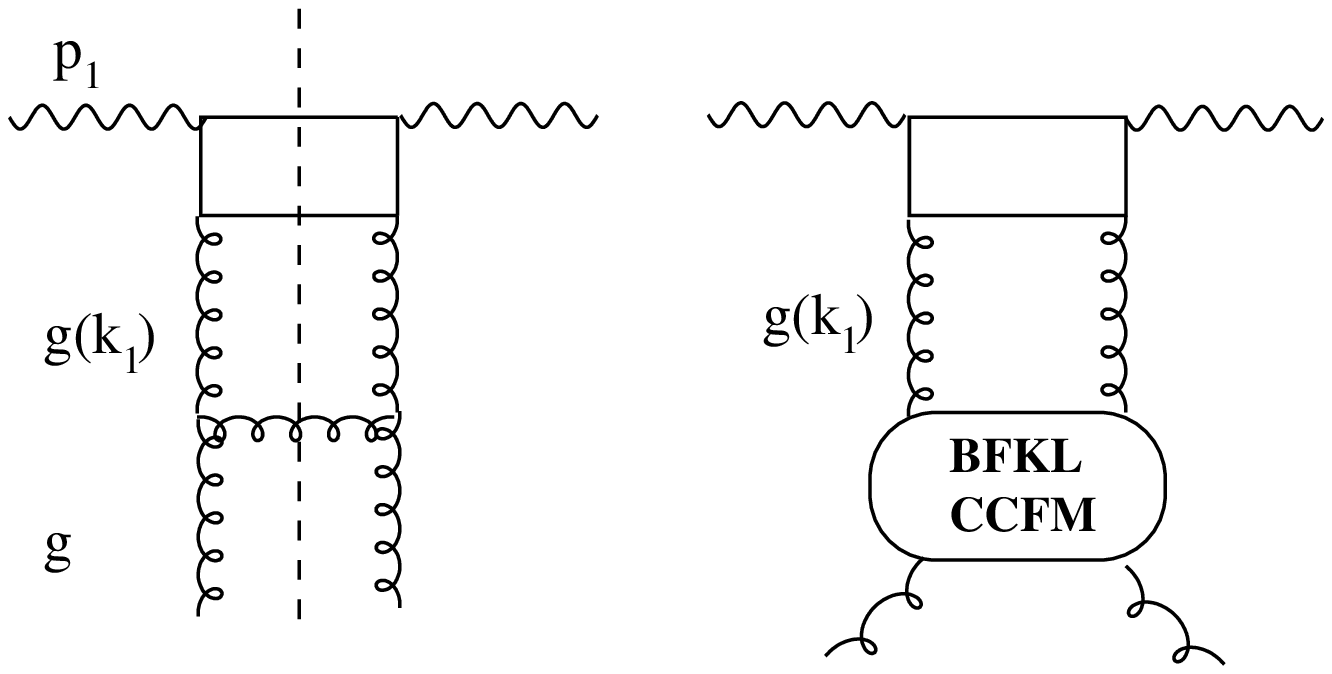,width=8cm}
\vskip -2cm
    \caption{{\it Schematic diagrams for $k_t$-factorization: 
    left is shown
    the one-loop correction to the Born diagram for photo-production
    right is shown the all-loop improved correction with the 
    factorized structure
    function ${\cal F}(x,\kt^2)$
    }}
   \label{fig:kt-fact}
  \end{center}
  \vspace*{-0.5cm}
\end{minipage}
\hskip 1cm
\begin{minipage}{0.45\textwidth}
\begin{center}
\vskip -0.7cm
\vskip -0.3cm
\epsfig{figure=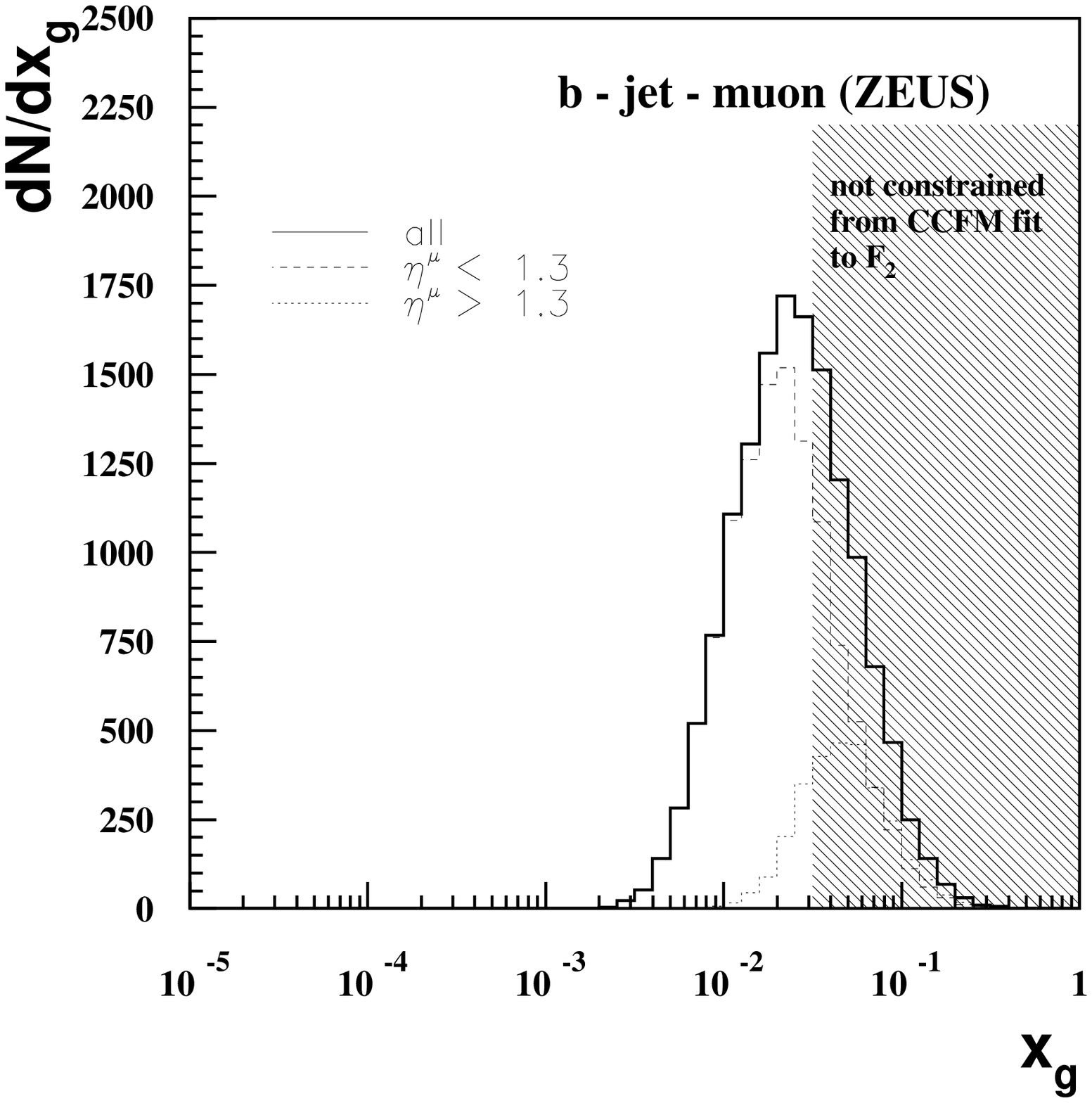,width=5.5cm,height=5.5cm}
\vskip -0.5cm
\caption{{\it The fractional energy $x_g$ of the gluon of the process $\gamma
g^* \to b \bar{b} $ satisfying the kinematic selection of 
ZEUS~\protect\cite{ZEUS-eps496}.
}}\label{zeus_xgluon}
\end{center}
  \vspace*{-0.5cm}
\end{minipage}
\end{figure}

The ${\cal O}(\as)$ matrix element in $k_t$-factorization includes 
the ${\cal O}(\as)$ matrix element of collinear factorization but in addition
also higher order contributions because 
the incoming gluon is off-shell and the
unintegrated gluon density resums    
parts of the virtual corrections (Fig.~\ref{fig:kt-fact}).
\par
The unintegrated gluon density 
$x {\cal A}(x,k_{t}^2,\Pmax)$\cite{jung_salam_2000,jung_ringberg2001,jung-hq-2001}, 
was obtained in the framework of
$k_t$ factorization using the CCFM evolution equation to 
fit\cite{jung_salam_2000} the structure
function $F_2(x,Q^2)$. This gluon density $x {\cal A}(x,k_{t}^2,\Pmax)$
is used in the hadron level Monte Carlo generator
\CASCADE \cite{CASCADEMC} for any comparison with measurements.

\section{Heavy quark production at HERA}
The prediction of \CASCADE\
for the total $b\bar{b}$ cross section\cite{jung_salam_2000}
 was compared to
the extrapolated measurements of the
H1\cite{H1_bbar} and ZEUS\cite{ZEUS_bbar} experiments at HERA. 
Since \CASCADE\ generates full hadron-level events, 
a direct comparison with measurements can be done,
before extrapolating the measurement over the full phase space 
to the total $b\bar{b}$ cross section. It has been  
shown\cite{jung_ringberg2001}, that
the visible $b\bar{b}$ cross section as measured by the 
ZEUS experiment\cite{ZEUS_bbar} is in perfect agreement 
with the predictions from \CASCADE . However, large uncertainties come from the
extrapolation of the measured cross section to the total $b\bar{b}$ cross
section. If the extrapolation is performed 
with \CASCADE\ (or any other program of the Lund family), 
approximate agreement with the NLO prediction is  also achieved 
in the case of the ZEUS measurement\cite{ZEUS_bbar}. 
\par
The prediction of \CASCADE\  has also been compared\cite{jung_ringberg2001}
to the measurement of H1\cite{H1_bbar}, where
it was also shown that the experimental results from H1 and ZEUS are
different by a factor $\sim 2$. 
It has been checked, that the difference between the experiments is not due to
different kinematics.

\begin{figure}[htb]
\vspace*{-7mm}
\begin{minipage}{0.45\textwidth}
\begin{center}
\epsfig{figure=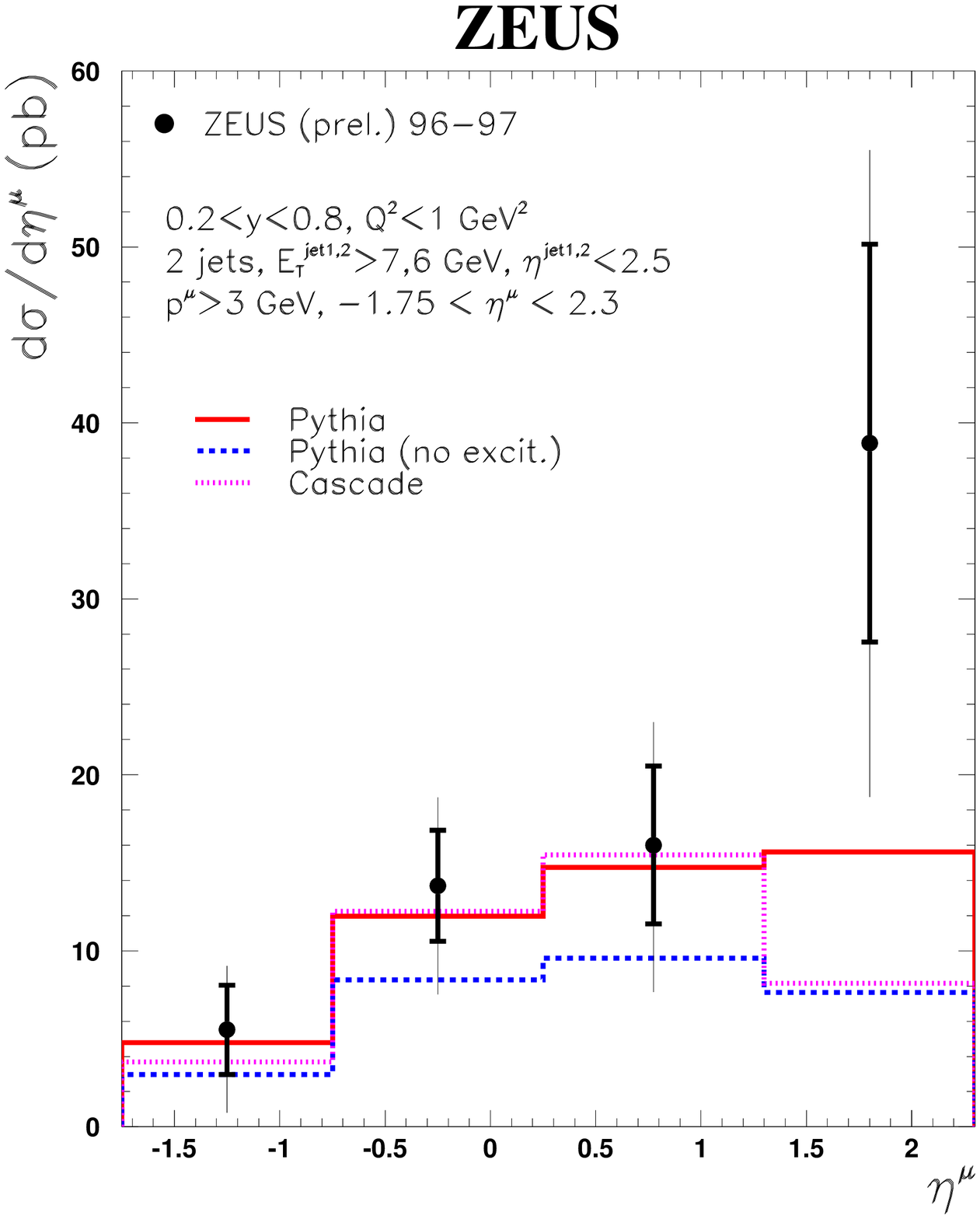,width=6cm,height=6cm}
\end{center}
\end{minipage}
\hspace*{0.6cm}
\begin{minipage}{0.45\textwidth}
\begin{center}
\epsfig{figure=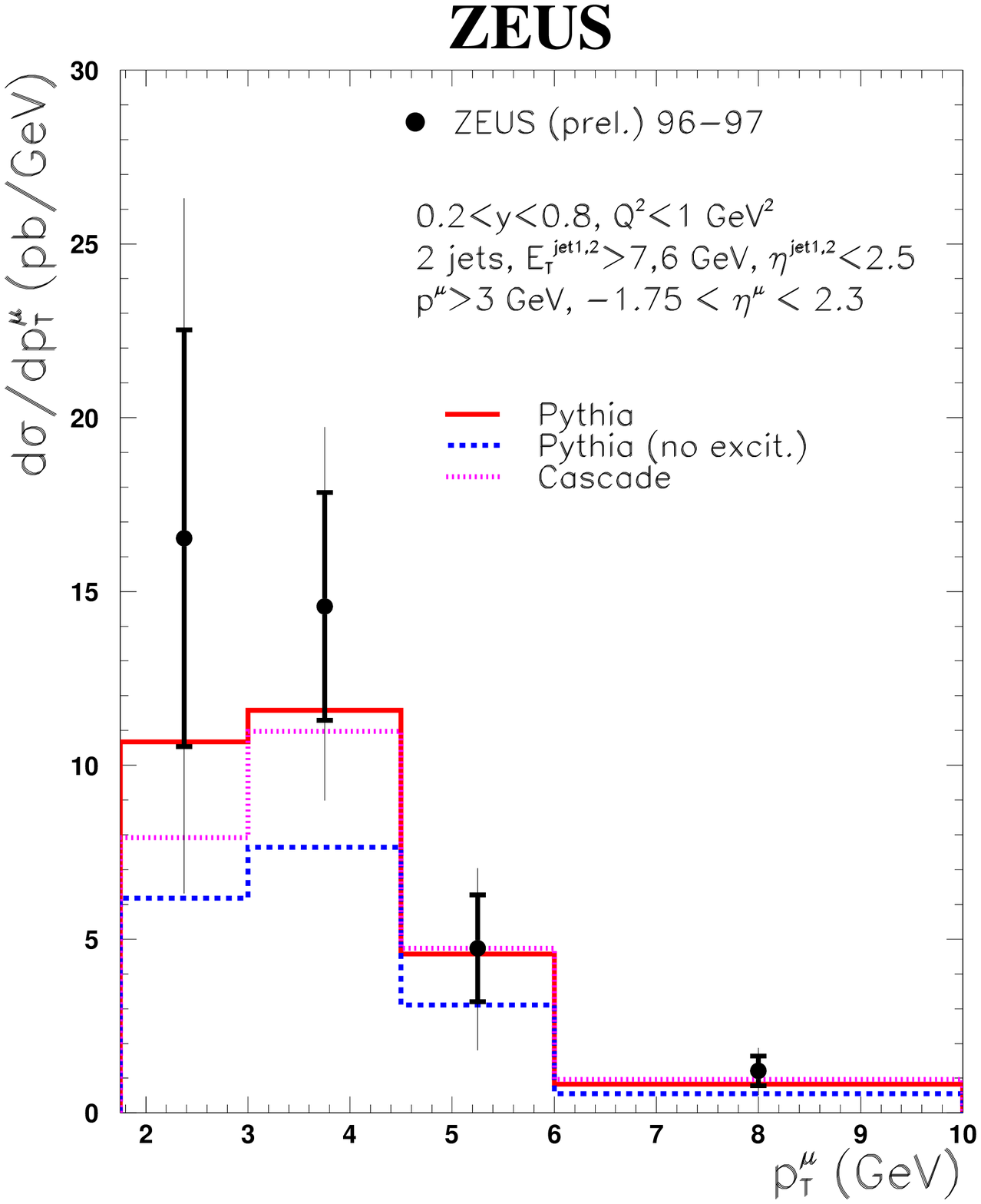,width=6cm,height=6cm}
\end{center}
\end{minipage}
\vskip -0.3cm
\caption{{\it Differential bottom cross 
section~\protect\cite{ZEUS-eps496} as a function of the muon
pseudo-rapidity (transverse momentum) for event with two jets and a muon,
compared to \CASCADE\ and \PYTHIA\ (with and w/o heavy quark excitation).
}}\label{zeus_muon}
\vspace*{-5mm}
\end{figure}
\par
In Fig.~\ref{zeus_muon} the differential cross section of $\mu$'s coming from
$b$-quark decays as measured by ZEUS\cite{ZEUS-eps496} is compared with the
prediction of \CASCADE . Again good agreement is observed, except in the region
of large $\eta^{\mu}$. However this region is dominated by large 
$x_g \gsim 0.03$ (see Fig.~\ref{zeus_xgluon}), where the application of 
$k_t$-factorization is questionable and also where the unintegrated
gluon density is not at all constrained by the fit to $F_2$. 
\begin{figure}[htb]
\vspace*{-2mm}
\begin{minipage}{0.45\textwidth}
\begin{center}
\vskip -0.7cm
\epsfig{figure=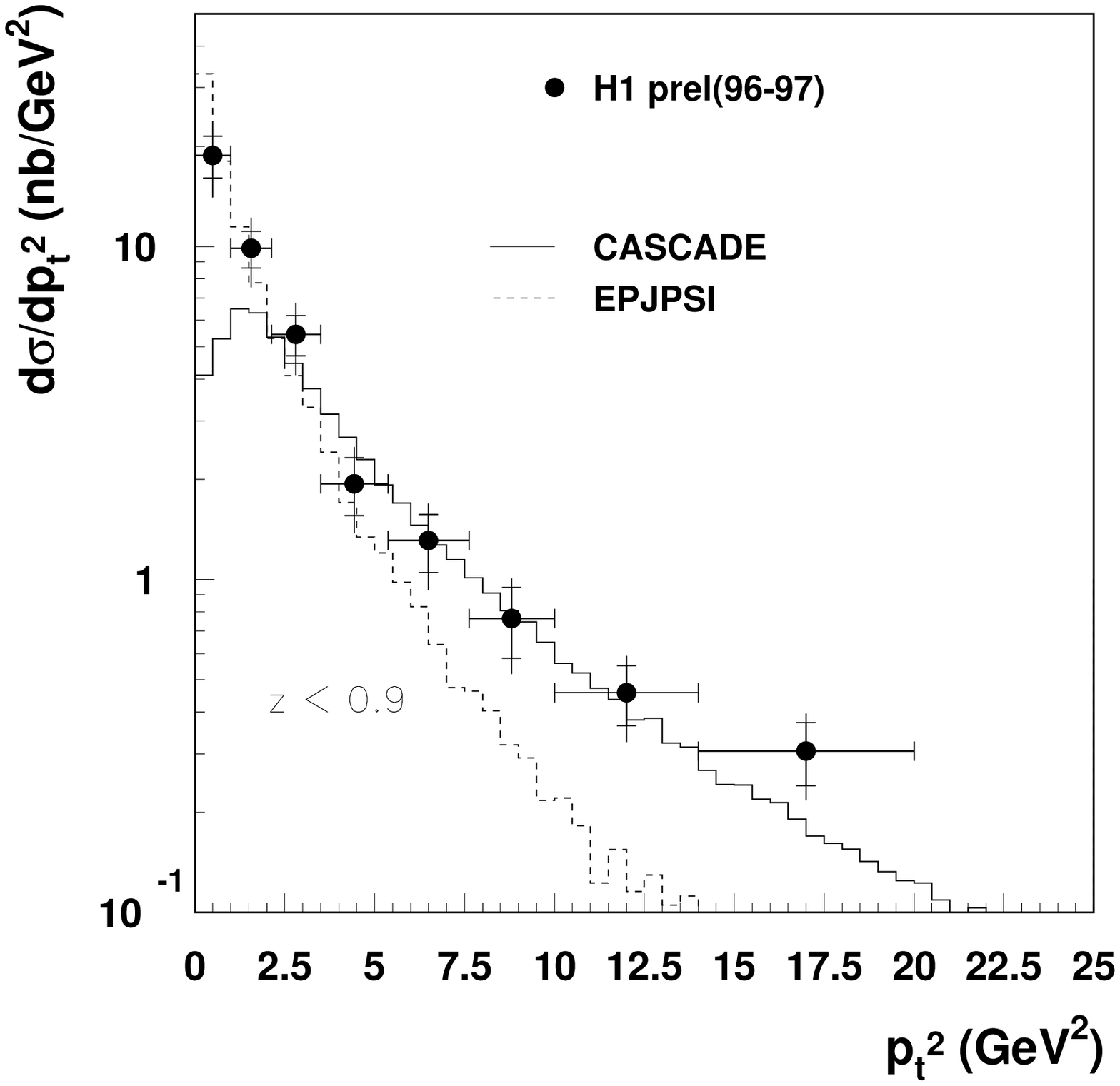,width=6cm,height=6cm}
\vskip -0.5cm
\end{center}
\end{minipage}
\hspace*{0.7cm}
\begin{minipage}{0.45\textwidth}
\begin{center}
\vskip -0.7cm
\epsfig{figure=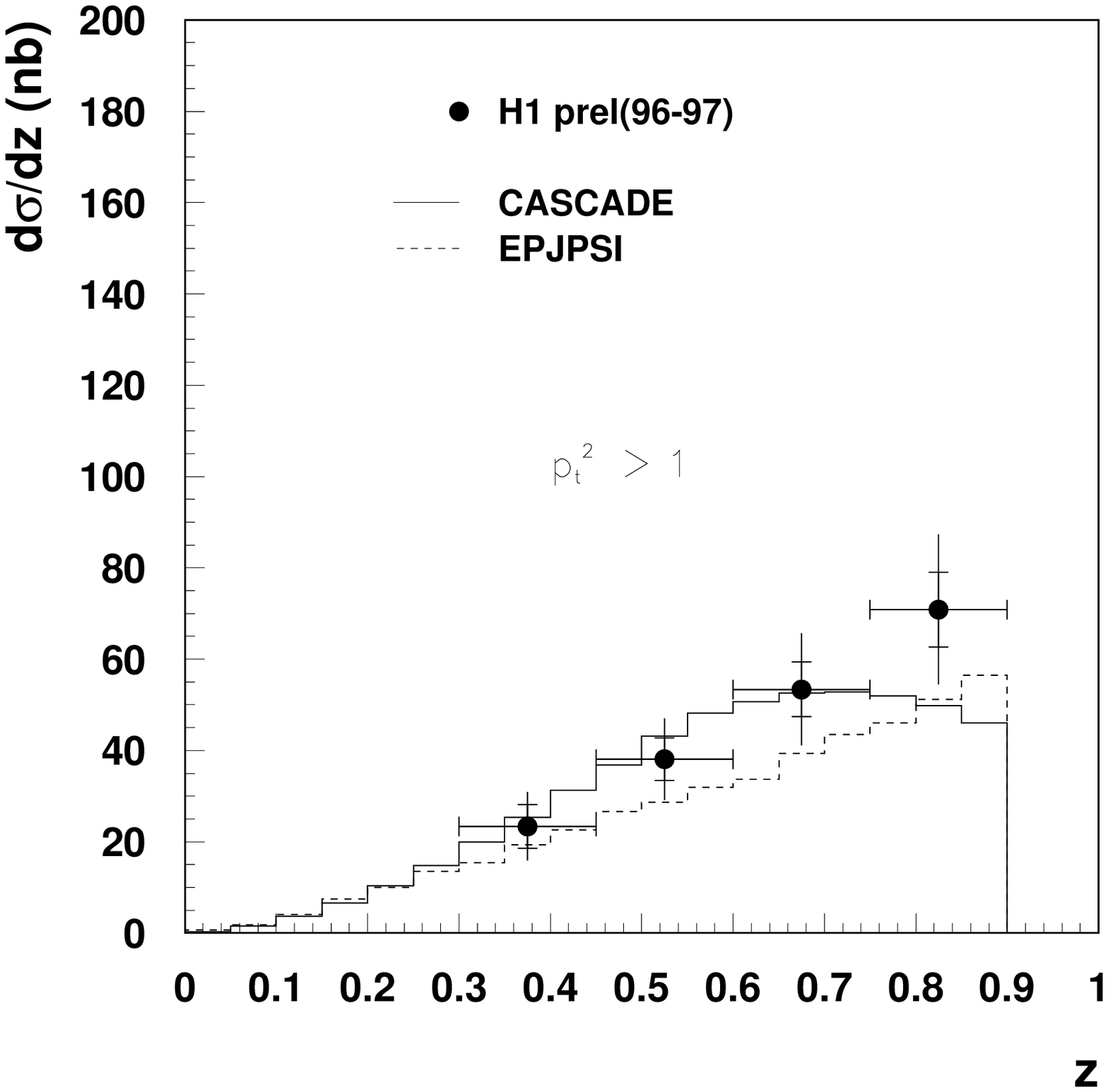,width=6cm,height=6cm}
\vskip -0.5cm
\end{center}
\end{minipage}
\vskip -4.6cm 
\hspace*{4.5cm}($a$)\hspace*{6cm}($b$)
\vskip 4.2cm 
\caption{{\it The cross section of $\gamma p \to J/\psi + X$
as a function of the transverse momentum ($a$)  
and of of the fractional momentum $z$ ($b$) of the $J/\psi$ as measured by 
H1~\protect\cite{h1_jpsi_99} compared to the predictions of \CASCADE\ and
\EPJPSI .
}}\label{h1_jpsi}
\vspace*{-5mm}
\end{figure}
\par
The  cross section for $\gamma p \to J/\psi + X$  
as a function of the transverse
momentum of the $J/\psi$ shows a significant deviation from the leading order
color singlet model prediction (as implemented in \EPJPSI\ \cite{EPJPSI3})
in collinear factorization.
In Fig.~\ref{h1_jpsi}, the
preliminary H1 measurement\cite{h1_jpsi_99}
 is compared to the prediction of 
\EPJPSI\ \cite{EPJPSI3}. In
collinear factorization, the harder transverse momentum spectrum is interpreted
as a signal for significant next-to-leading order corrections. Also shown in 
Fig.~\ref{h1_jpsi}$a$ is the prediction of \CASCADE\ using the same CCFM
unintegrated gluon density as before together with the $k_t$-factorized matrix
element\cite{saleev_zotov_a} for $\gamma g^* \to J/\psi g$. The
inelastic $J/\psi$ photoproduction cross section ($z<0.9$) as a function
of $p_T$ is nicely described by \CASCADE .
 Especially the large transverse momentum part is explained
as additional hard initial state QCD radiation. The same is also observed in
full NLO calculations, and it shows again the advantage
of the $k_t$-factorization in simulating 
a large part of the NLO correction of the collinear approach, due to
the non zero virtuality of the incoming gluon. The distribution in the
fractional momentum $z$ of the $J/\psi$  is also reasonably well described 
(Fig.~\ref{h1_jpsi}$b$).
\section{Conclusion}
The application of the $k_t$-factorization approach to heavy quark production
at HERA has been discussed. It is shown, that the visible 
$b\bar{b}$  production cross section
as measured at ZEUS is nicely reproduced by \CASCADE , using
the CCFM evolved unintegrated gluon density obtained from a fit to 
$F_2(x,Q^2)$. 
It was pointed out, that the extrapolation from the measured to the total
$b\bar{b}$ cross section contains large model dependencies.
\par
The differential  cross sections for inelastic
$J/\psi$ photoproduction can be reasonably well described with \CASCADE , both
in terms of shape and normalization. The large transverse momentum tail is a
direct signal for additional hard initial state QCD radiation.
\par
It is the advantage of the $k_t$-factorization approach
that important parts of NLO and even NNLO
contributions are consistently included due to the off-shell gluons, which
enter into the hard scattering process.   
\section*{Acknowledgments}
I want to thank the organizers  M.~Kienzle and M.~Wadhwa for this very 
nice workshop. I am also grateful to S. Frixione for the invitation to 
this workshop and his interest in $k_t$-factorization.
Many thanks also go to S.~Baranov and N.~Zotov for interesting discussions
and our fruitful collaboration.
I am also grateful to  L. Gladilin, B. Naroska and J. Whitmore for careful
reading of the manuscript.
All thanks for the great times with Antje.

\end{document}